\documentclass[prl,aps,twocolumn,showpacs]{revtex4}

\usepackage{dcolumn}
\usepackage{amssymb}
\usepackage{amsmath}
\usepackage{graphicx}
\usepackage{epsfig}
\usepackage{color}

\newcommand \be {\begin{equation}}
\newcommand \ee {\end{equation}}
\newcommand \bea {\begin{eqnarray}}
\newcommand \eea {\end{eqnarray}}

\renewcommand{\vec}[1]{{\bf #1}}

\begin{document}

\title{Connecting  diffusion and dynamical  heterogeneities in actively deformed amorphous systems}

\author{Kirsten Martens}
\affiliation{LPMCN, Universit\'e de Lyon; UMR 5586 Universit\'e Lyon 1 et CNRS, F-69622 Villeurbanne, France}

\author{Lyd\'eric Bocquet}
\affiliation{LPMCN, Universit\'e de Lyon; UMR 5586 Universit\'e Lyon 1 et CNRS, F-69622 Villeurbanne, France}

\author{Jean-Louis Barrat}
\affiliation{LIPHY, Universit\'e Grenoble 1 et CNRS, F-38402 Saint Martin d'H\'eres, France}

\date{\today}

\begin{abstract}
In this letter we explore the relations between tracer diffusion and flow heterogeneities in flowing
amorphous materials. 
On the basis of scaling arguments as well as an extensive numerical study of an athermal elasto-plastic model, 
we show that there is a direct link between
the self-diffusion coefficient and the size of cooperative regions at low strain rates. Both depend strongly on 
rate and on system size. A measure of the mean square displacement of passive tracers 
in deformed amorphous media thus gives information about the microscopic rheology, such as
the geometry of the cooperative regions and their scaling with strain rate and system size. 
\end{abstract}


\maketitle

Under the application of sufficiently large stresses,  
solid amorphous materials, such as dense emulsions \cite{Goyon}, colloidal and granular systems  \cite{Candelier10Schall10} or molecular glasses \cite{Ediger09}, undergo plastic deformation and flow. 
This yielding behavior usually comes along with peculiar spatial features, such as deformation heterogeneities, shear bands \cite{SB} or nonlocal flow behavior \cite{Goyon}. At the fundamental level, the emerging picture to describe these properties
involves local dissipative rearrangements, the so called 'shear transformations' introduced by Argon \cite{Argon1979}, 
cascading via long range elastic interactions. 
This scenario has been largely confirmed by a number of numerical
\cite{FalkLanger1998,MaloneyTanguy} and experimental \cite{Schall07} studies of quasistatic deformation in amorphous systems. 
It was also fruitful on the theoretical side, and served as the basis for various rheological descriptions, which aim at describing the statistical collective properties
of plastic events \cite{FalkLanger1998,SGR,BocquetFieldingLanger}. Yielding, nonlocality in the flow behavior, enhanced mobility
\cite{Ediger09} and heterogeneities are therefore only different aspects of the same underlying physics.



\begin{figure}[h]
\centering\includegraphics[width=0.6\columnwidth,clip]{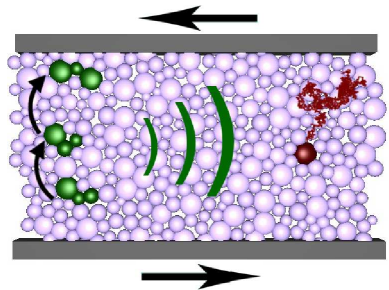}
\caption{(color online) Schematic view on the heterogeneously distributed cascading plastic events (on the left) and the resulting long range effects on the diffusion of a tracer particle (on the right). }
\label{figure1}
\end{figure}

In this letter, we consider the question of tracer diffusion in such flowing materials. 
On the experimental side, results indicate that the diffusion of tracer particles, e.g. in sheared foam systems and in colloidal
 glasses, is strain rate dependent \cite{vanHecke10b, Weeks07}. Further observations concern the strong nonlocal effects 
 in the mechanical response even in quiescent regions far away from the flowing material \cite{vanHecke10a}. 
%
 The observation of a coupling between enhanced particle diffusion and an imposed deformation is quite natural in the context of 
 slowly driven systems,
 where the external driving provides a source of energy that may share some characteristics of a thermal bath \cite{CKP-BBK}. 
Here, we investigate this connection based on the specific deformation mechanisms associated with the flow behavior. Tracer diffusion 
is expected to take its origin in the mechanical noise induced by the spatio-temporal collective behavior of local plastic events. 
In this context, we first establish 
a direct link between flow induced dynamical heterogeneities and the diffusive dynamics of tracer particles 
immersed in a deformed system (see Fig.~\ref{figure1}). We summarize this in the form of a general scaling relation between the diffusion coefficient and the 4-point correlations that quantify flow heterogeneities. This allows us to rationalize the strain rate dependence, as well as a strong finite-size dependence of the diffusion coefficient in sheared amorphous materials. We first discuss the general scaling arguments for such a connection, and in a second part we support these arguments on the basis of large scale simulations of an elasto-plastic model of yield stress materials.

{\it A scaling argument.} --
 We start from the microscopic picture of 'shear transformation events'. Each event  is assumed to correspond to a local Eshelby  transformation \cite{Picard}, 
with a strain amplitude $\Delta\varepsilon_0$  localized in a zone of linear size $a$. The displacement field for an event occurring at the origin 
decays algebraically; in 2D it reads $\vec{u}(\vec{r})=({2 a^2} \Delta \epsilon_0{xy}/\pi r^4)\times \vec{r}$.
 If we consider the non-affine part of the dynamics of tracer particles in an athermal medium, their relative position $\vec{r}_T$ will be determined by the time integrated velocity field $\vec{v}$ caused by the surrounding plastic events,
\begin{equation}
\vec{r}_T(t)=\vec{r}_T^{(0)}+\int_0^t dt'\sum_{i=0}^{N_p(t')} \vec{v}(\vec{r}_T(t')-\vec{r}_i(t'))
\end{equation}
with $\vec{r}_T^{(0)}=\vec{r}(t=0)$, $\vec{r}_i$ the position of the $i$th plastic event and $N_p(t)$ the number of plastic events at time $t$.

To give an estimate for the mean square displacement of these tracer particles in the highly cooperative regime, we 
introduce the notion of cooperative events that span a cooperative 
 volume $V_c$, and have a typical duration $t^*$. The typical mean square displacement  due to an cooperative event (avalanche, slip-line or more complex object, depending on the geometry) can  be estimated as
$\langle\Delta l^2\rangle=\rho^2\int_{V_c}d^dr\int_{V_c}d^dr' C(\vec{r}-\vec{r'})$, 
with $\rho$ the density of elementary events in the cooperative volume and $C(\vec{r}-\vec{r'})=\langle \vec{u}(\vec{r})\vec{u}(\vec{r'})\rangle$.
This yields
\begin{equation}
\langle\Delta l^2\rangle\approx\frac{a^4\Delta\varepsilon_0^2}{\pi L^2}\rho^2 V_c^2(t^*) 
\end{equation}
up to logarithmic corrections originating from the displacement field correlations \cite{Lemaitre09}.
The number of the cooperative events, $N_c$, occurring during a strain interval $\Delta\gamma$ can be estimated by the total number of plastic events $N_p(\Delta \gamma)=L^2\Delta\gamma/a^2\Delta \varepsilon_0$ divided by the number of events in the cooperative volume, 
$N_c(\Delta \gamma)=N_p/\rho V_c$. If we further assume that the cooperative events are statistically independent, the diffusion coefficient $D$ for the tracer particles in the limit of weak strain
rate $\dot{\gamma}$ can be written as $D=\dot{\gamma}\tilde{D}$ with
\begin{equation}
\tilde{D}=\frac{N_c \langle\Delta l^2\rangle}{2\Delta\gamma}\approx\frac{a^2\Delta\varepsilon_0}{2\pi}\rho V_c\;.
\label{main-result}
\end{equation}
The appropriate tools for describing cooperativity are the '4-point' correlation functions, that quantify
the  correlation in space and time between events, and are described in more detail below. The 4-point function of interest here will be the one 
associated with plastic activity, rather than particle displacements. The integral over space of such correlation functions,
denoted by $\chi_4(t)$
goes in general trough a maximum at a finite time $t^*$, which can be used as a measure of the volume over which 
events are correlated.  Therefore we expect, in macroscopic systems 
that the diffusion coefficient should scale in the same manner, i.e.
\begin{equation}
\tilde{D}\sim \chi_4(t^*)\;.
\label{D-chi4}
\end{equation}
 As we will demonstrate in the example below  $\chi_4(t^*)$ is not only strain rate but also strongly system size dependent.
 This behavior originates from the long range character of the displacement field created by the plastic events. 
 As a result we recover the same strong nonlocal effects for the diffusion coefficient. This is consistent with the experimental observations, that predict a sensitive mechanical response to distant active regions \cite{vanHecke10a}, and with results from simulations of simple particle systems \cite{Lemaitre09}. 

{\it An elasto-plastic mesoscale model} --
To test the ideas described above, we performed large scale simulations \cite{tobepublished} of a  simple mesoscopic model for the flow of yield stress materials,  introduced by Picard et.~al.~\cite{Picard}. The model  is restricted to athermal systems and to a plane geometry. It features the dissipation due to local plastic events and the associated elastic response, that relaxes the stresses over the system. These minimal ingredients are sufficient to generate a complex rheological behavior
\cite{Picard}. Approximating the rearrangements by spherical inclusions (Eshelby problem), one expects a four-fold quadrupolar symmetry for the inhomogeneous part of the stress propagator $G(r,\theta)=\frac{1}{\pi r^2}\cos(4\theta)$. 
To simplify  the model further we consider an incompressible medium and assume that the microscopic geometry of the plastic events is the same as the the one of the macroscopic shear, which permits a  scalar description of the stress. The values of this local stress $\sigma_{i\alpha}$ are encoded on a square lattice of size $N=L^2$, given in units of $a^2$, the typical size of a plastic event (e.g. several grains in case of granular material).

The deterministic part of the stress dynamics reads, using  dimensionless  quantities
\bea
\label{det-eq}
\partial_t \sigma_{i\alpha}(t)&=&\dot{\gamma}+2\sum_{j=1}^{L_x}\sum_{\beta=1}^{L_y} G_{\alpha\beta}^{ij}\dot{\varepsilon}_{j\beta}^{pl}
\eea
where the time $t$ is measured in units of an elementary  relaxation time $\tau$, the stress $\sigma$ in units of the local yield stress $\sigma_y$ and the strain
 rate $\dot{\gamma}$ in units of the critical  value $\dot{\gamma}_c=\sigma_y/\tau\mu$, $\mu$ being the shear modulus.  $\dot{\gamma}_c$ indicates the change from a Newtonian to a non Newtonian flow behavior.  $G_{\alpha\beta}^{ij}$ is the discretized propagator for a finite geometry with periodic boundary conditions \cite{Picard}. We assume for the dynamics of the plastic part of the strain $\dot{\varepsilon}^{pl}$ a viscoelastic like relaxation of the material, weighted with a local state variable $n_{i\alpha}$ indicating whether a site is plastically active or not
$\dot{\epsilon}_{i\alpha}^{pl}=n_{i\alpha}(t)\sigma_{i\alpha}(t)/2$. 
The stochastic dynamics of the activity is ruled by the following dynamics for the plastic state variable 
$$n_{i\alpha}: \quad 0\xrightarrow[\sigma_{i\alpha}>1]{\tau_\mathrm{plast}^{-1}} 1\quad  0\xleftarrow[\forall \sigma_{i\alpha}]{\tau_\mathrm{elast}^{-1}} 1 \;.$$
In simulations we chose $\tau_\mathrm{plast}=\tau_\mathrm{elast}=1$ in units of $\tau$. Note that convection effects on the stress and the activity have been neglected to keep the model as simple as possible. The lack of convection leads basically to an additional symmetry for the spatial arrangements of the plastic events. This is expected to affect the shape
of the heterogeneities, but will not alter the general relation between dynamical heterogeneities and the diffusion dynamics.

\begin{figure}[t]
\centering\includegraphics[width=\columnwidth,clip]{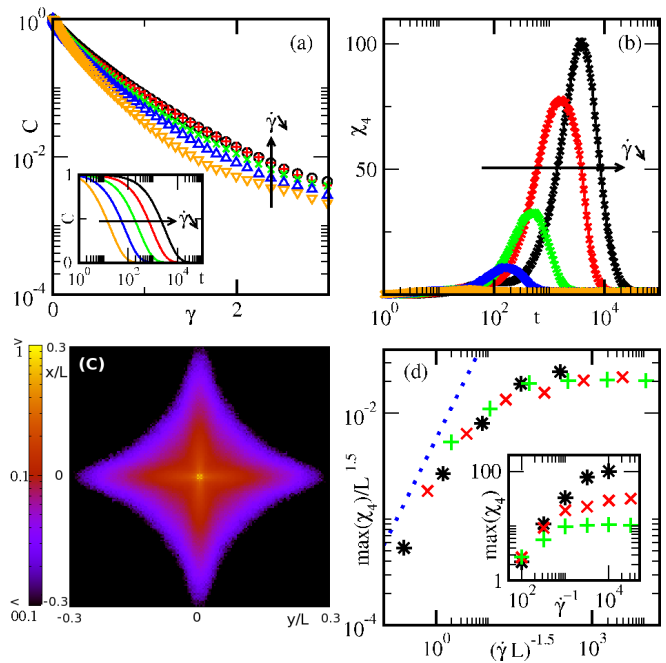}
\caption{(color online). Dynamical heterogeneities in the stress field. 
 When not otherwise specified, the system size is $N=2^{16}$. 
 {\it Top left}: Normalized two-time auto-correlation function of the stress as a function of strain ($\dot{\gamma}=10^{-2},10^{-2.5},10^{-3},10^{-3.5},10^{-4}$). The inset shows the same quantity as a function of time. {\it Top right}: Time evolution of the dynamical susceptibility (same values for $\dot\gamma$). {\it Bottom left}: Spatial shape of the normalized four point stress correlation function $G_{4}$ for a strain rate $\dot{\gamma}=10^{-4}$ at time $t^*$ where the dynamical susceptibility is maximal. {\it Bottom right}: Finite size scaling plot for the maxima of the dynamical susceptibilities. Shown are the maxima as a function of the inverse strain rate, both normalized by the system size ({\color{green} +} $N=2^{12}$, {\color{red} $\times$} $N=2^{14}$, $\ast$ $N=2^{16}$). The dotted line has slope one and the inset shows the raw data.}
\label{figure2}
\end{figure}

{\it Dynamical heterogeneities.} --
To obtain the scaling of the cooperative volume $V_c$ entering  equation ~\ref{main-result}, we study the dynamical correlations within the model described above. Under the influence of the external strain rate $\dot{\gamma}$, the system reaches a non equilibrium steady state, where two time correlations of all quantities in the system become time-translation invariant. The time needed to achieve this steady state increases as the inverse of the strain rate.
A first indication for growing dynamical heterogeneities with decreasing strain rate is the behavior of the  two time correlation function of the local stress fluctuations $\rho_{i\alpha}(t)=\sigma_{i\alpha}-\sigma_d$, where $\sigma_d$ is the steady state stress averaged over time and space. The correlation function  reads
$C(t)=\langle\overline{\rho_{i\alpha}(0)\rho_{i\alpha}(t)/\rho(0)^2}\rangle_c$, 
where the bar indicates a spatial average and the brackets a configuration average. In the limit of low strain rate, a data collapse is obtained when $C$
 is plotted as a function of strain (see Fig.~\ref{figure2}a). The relaxation time is therefore inversely proportional to strain rate, although a deviation from this behavior is observed as the strain rate increases.
 A hint to the existence of a complex, cooperative dynamics lies in the non exponential behavior of $C$, which can be interpreted as resulting from the coexistence of regions with fast and slow dynamics.
 Such dynamical heterogeneities are revealed by the study of a four point correlation function, 
\bea
G_4^{(i\alpha)} (t)&=& \sum_{j\beta}\left[\langle\rho_{i\alpha}(0)\rho_{i\alpha}(t)\rho_{j\beta}(0)\rho_{j\beta}(t)\rangle_c\right.\nonumber\\
&&-\left.\langle\rho_{i\alpha}(0)\rho_{i\alpha}(t)\rangle_c\langle\rho_{j\beta}(0)\rho_{j\beta}(t)\rangle_c\right]\;.
\eea
The integral of $G_4$ over space yields the variance of the two time correlation function, 
 $\chi_4(t)=N (C_2(t)-C^2(t))$ with $C_2(t)=\langle\overline{(\rho_{i\alpha}(0)\rho_{i\alpha}(t)/\rho(0)^2)}^2\rangle_c$.  $\chi_4$ generically displays 
 a maximum at a time $t^*$ corresponding to the time of the largest heterogeneity of the system, 
 and in general is the typical decay time for $C$. Moreover, $\chi_4(t^*)$ can be interpreted as the number of 
 sites involved in events at the cooperativity maximum, and is therefore identified with the number $V_c$ introduced in the first part.
 
 As expected from the study of $C$, $t^*$ scales with the inverse of the applied strain rate. A more interesting feature is the growth of $\chi_4(t^*)$ as $\dot\gamma\rightarrow 0$, which clearly indicates the increasingly heterogeneous behavior of the system.
A more detailed study of $G_4$ shows that this function is strongly anisotropic, a result of the special form of the stress propagator (see Fig.~\ref{figure2}c). Integrating $G_4$ over space reveals a fractal dimension of the cooperative domains close to 3/2, which is consistent with the finite size scaling displayed in  Fig.~\ref{figure2}d. 

To obtain this scaling we plot the height of the peak as a function of $\dot{\gamma}$ for different system sizes in the inset of Fig.~\ref{figure2}d. Clearly we observe a strong size dependence. In an infinite system we expect that the fractal dynamical cooperation length, traditionally denoted as $\xi_4(\dot{\gamma})$,  would grow as a negative power of $\dot{\gamma}$. However, in a finite system, this length will saturate when reaching the limit of the sample. This effect leads to the plateau region in the $\chi_4$ curve with value $\chi_0$. For a fixed system size we reach this plateau below a critical strain rate $\dot{\gamma}_{\chi}$. It is then
natural to write the following finite-size scaling
\begin{equation}
\chi_4(\dot{\gamma},L)=L^{3/2} f\left(\left(\xi_4(\dot{\gamma})/L \right)^{3/2}\right)
\end{equation}
with a scaling function $f(x)$, that is linear for small values of $x$ and saturating for large $x$. This hypothesis is perfectly confirmed in Fig.~\ref{figure2}d, with $\xi_4\sim\dot{\gamma}^{-1}$.
We expect the specific value for the fractal dimension 
for the geometry of the heterogeneities to depend on some specific details of the model. Here, cooperative events are free to spread in the directions of the two main axes due to the additional symmetry introduced through the absence of convection in the model system. This results into fractal structures with dimension larger than unity, rather than one-dimensional slip lines. We believe this behavior to be representative of three dimensional systems where an additional direction is introduced, in agreement with earlier results on 3D molecular dynamics simulations \cite{Bailey07}.

{\it Diffusion} --  
Within the present model, we introduce diffusion of tracers by considering the non-affine motion originating 
from the long range displacement fields induced by  plastic events. We then  associate with each (mesoscopic) event a corresponding continuous displacement field (as introduced in the former scaling argument) 
and use it to define the mean square displacement of the tracer particles. 
Noninteracting tracer particles are assigned to every lattice site, and their fictitious trajectory is built progressively by adding up contributions from all plastic events. In this way, the mean square displacement $\langle \Delta r^2\rangle$ can be obtained and a diffusion coefficient $\mathcal{D}=\dot{\gamma}\tilde{\mathcal{D}}$ with
\begin{figure}[t]
\centering\includegraphics[width=\columnwidth,clip]{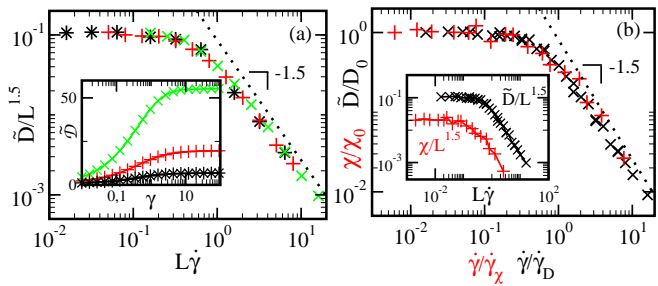}
\caption{(color online) {\it Left}: scaling of the diffusion coefficient $\tilde{D}$ with strain rate and system size  ($\ast$ $N=2^{8}$,  {\color{red} +} $N=2^{10}$, {\color{green} $\times$} $N=2^{12}$). The inset shows  $\tilde{\mathcal{D}}$ versus strain $\gamma$ for $\dot{\gamma}=10^{-3}$. {\it Right}: Master curves for diffusion coefficient ($\times$) and dynamical susceptibility ({\color{red} +}), normalized by their plateau value as a function of strain rate in units of the corresponding critical strain rate, indicating the onset of finite size effects. The inset shows the unscaled master curves.}
\label{figure3}
\end{figure}
$\tilde{\mathcal{D}}=\langle \Delta r^2\rangle/2\Delta\gamma$ can be extracted with a good accuracy. The mean square displacement as a function of strain  is ballistic for small strains,  and becomes diffusive after a transient regime ($\Delta{\gamma}\approx 1$), with a well defined  slope $\tilde{D}$, see inset in Fig.~\ref{figure3}a. This quantity varies strongly with strain rate (for strain rates $\dot{\gamma}<\dot{\gamma}_c$) and system size (for all $\dot{\gamma}$), see Fig.~\ref{figure3}a. For a given system size we find power law behavior for the scaling with strain rate $\tilde{D}\sim\dot{\gamma}^{-1.5}$ down to a critical  rate $\dot{\gamma}_D$ below which a plateau is reached. This plateau value $D_0$ scales with system size as $D_0\sim L^{-1.5}$. To summarize, we find the {\it same scaling behavior as for the dynamical susceptibility}, see Fig.~\ref{figure3}b, with a crossover  length scale for diffusion  $\ell_D$ proportional to $\dot{\gamma}^{-1}$.

Therefore we validate the scaling prediction in Eq.~\ref{D-chi4}, $\tilde{D}\sim \chi_4$, for large values of $L\dot{\gamma}$. 
Only the onset on the finite size effects occur for different parameter values, $\dot{\gamma}_D\approx 3\dot{\gamma}_{\chi}$ (see inset of Fig.~\ref{figure3}b). This indicates that the intrinsic length scales for the heterogeneities, $\xi_4$ and  for the diffusion, $\ell_D$, are not identical but only proportional with $\xi_4<\ell_D$. 

{\it Discussion.} --
Our analysis shows  how the mechanism of deformation in amorphous materials at low temperature, described by elastic interactions between 
isolated shear transformation events, results in nontrivial cooperative effects that make a strong contribution to passive tracer 
 diffusion at low strain rates. This connection is embodied in a scaling relation between the diffusion coefficient and the cooperative volume
 for plastic heterogeneities.
 We stress that the  diffusion considered here is associated with the long range part of the elastic deformation induced by plastic events, thus
 does not involve the more direct one,
 associated with the motion of particles at the core of the transformation zones. The former is expected
 to be dominant at small strain rate.   
 The peculiarity of this diffusion is a strong system size dependence, that reveals the non local aspect of the system dynamics. 

In the literature, several observation of strain rate and system size dependent diffusion constants have been reported, both in experiments \cite{Weeks07,Weitz10}
and simulations \cite{Lemaitre09,Tsamados10Heussinger10}. 
Our results suggest that in addition to strain rate effects,
the measured diffusion may be strongly affected by system size effects.   
It would therefore be desirable to perform experimental finite-size scaling analyses of tracer diffusion in flowing amorphous
materials,
using {\it e.g.} confined microfluidic systems, 
which provide key information about the rheological heterogeneities. 

{\it Acknowledgements} We acknowledge financial support from ANR, program SYSCOMM. KM was supported by 
the Swiss National Science Foundation and the Marie Curie 
program.  JLB is supported by IUF. 
LB thanks A. Ajdari and F. Krzakala for useful discussions at the early stages  of this work.

%
%
%
%


\begin{thebibliography}{99}
\bibitem{tobepublished}
The numerical solution of the deterministic and stochastic equations have been performed using GPU computing. Technical details will be published elsewhere.
\end{thebibliography}
\end{document}